\documentclass[aps,twocolumn,groupedaddress]{revtex4}

\begin{document}
\draft

\title{Tunnelling of entangled Kondo singlet in two-reservoir nanocontact systems under bias}
\author{Jongbae Hong$^{1}$}
\author{S.G. Chung$^2$}
\affiliation{$^1$Department of Physics, POSTECH and
Asia Pacific Center for Theoretical Physics, Pohang, Gyeongbuk 790-784, Korea\\
$^2$Department of Physics, Western Michigan University, Kalamazoo, MI 49008-5252, USA}
\date{\today}

\maketitle \narrowtext

{\bf Tunnelling conductances observed for mesoscopic Kondo systems exhibit a zero-bias peak
and two coherent side peaks. The former peak is usually understood as a Kondo effect and the latter side peak
is recently clarified as the effect of inter-reservoir coherence. However,
fitting the experimental $dI/dV$ line shapes, where $I$ and $V$ denote the current and bias voltage, respectively,
has not been performed theoretically. Here, we fit the entire line shape range of the tunnelling conductance observed
for a quantum dot, quantum point contact, and magnetized atom adsorbed on an insulating layer
covering a metallic substrate by studying the tunnelling of entangled Kondo singlet (EKS) formed in a two-reservoir
mesoscopic Kondo system. We also clarify the characteristic dynamics forming each coherent peak
in terms of the processes comprising spin exchange, singlet hopping, and singlet partner changing.
Tunnelling of entangled Kondo singlet can be applied to understanding the tunnelling
conductance observed for a sample with strong electron correlation.}

Kondo-involved mesoscopic systems include a quantum dot single-electron transistor (QDSET)$^{1,2}$, quantum point
contact (QPC)$^3$, and magnetized atom adsorbed on a metallic substrate$^4$. The observations of tunnelling
conductance for these systems were reported around the same time. Such mesoscopic Kondo systems have been further studied by applying a magnetic field$^{5,6}$, varying the gate voltage$^{7,8}$, and covering a metallic substrate with an insulating layer$^{9,10}$. The tunnelling conductances observed from these systems are characterized by a zero-bias peak and two coherent side peaks, except for the adsorbed magnetized atom without an
insulating layer in which the zero-bias peak is suppressed$^4$.
We study the Kondo peak suppression in a separate work.

Interestingly, the central peak of the QDSET seems to be a single peak$^2$. However, we show
in the following text that the single peak comprises a zero-bias peak
and two side peaks. Hence, the mesoscopic Kondo systems under consideration in this study
have three coherent peaks, i.e., a zero-bias peak and two side peaks.
Legitimate questions are then, ``What is the origin of two side peaks?" and ``Why do the previous
approaches using the Keldysh formalism$^{11,12}$, the real-time renormalization group method$^{13}$,
quantum Monte Carlo calculations$^{14}$, and the scattering-state numerical renormalization group method$^{15}$
not show the side peaks?".
One of us (JH) answered for these questions, and the answer was inter-reservoir coherence$^{16}$.
The two side peaks signify the effect of inter-reservoir coherence and the previous studies
do not take it into account.

We study the two-reservoir Anderson impurity model in a fully coherent state,
in which the Kondo singlet is entangled, as shown in Fig.~1a. The left and right Kondo clouds
in Fig.~1a are within the coherent region. The crucial property of EKS tunnelling is moving together
unidirectionally similar to the superconducting Josephson current.
Backward scattering is allowed for a non-EKS which is formed by breaking the inter-reservoir
coherence. Our previous study$^{16}$ shows that the spectral function exhibiting
a bias-dependent split Kondo peak$^{11-15}$ is a characteristic of non-EKS state.

\begin{figure}
[b] \vspace*{4.0cm} \includegraphics{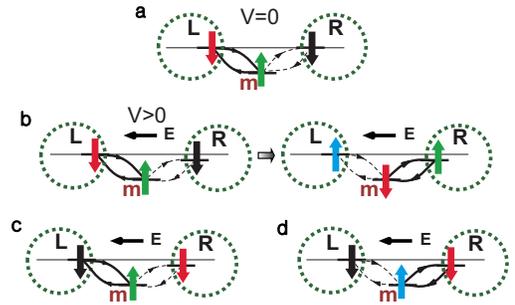} \vspace*{0.0cm}
\caption{\textbf{Spin dynamics in an entangled Kondo singlet.} {\bf a,} Entangled Kondo singlet at equilibrium.
{\bf b,} Spin configurations in singlet hopping. {\bf c,} A spin state after exchange and partner change
from the right configuration of {\bf b}. {\bf d,} A spin state after partner change and singlet hopping
from the right configuration of {\bf b}. The dashed circle, ``m", and ``E" denote the Kondo cloud,
mediating Kondo atom, and electric field, respectively. }
\end{figure}

Our main purpose is to obtain the $dI/dV$ line shapes which fit the entire
range of tunnelling conductances observed for the abovementioned mesoscopic Kondo systems.
A fitting as provided in this study has never been reported previously.
It is obvious that the line shapes of tunnelling conductance observed by two-terminal
experiments$^{2-10}$ for various mesoscopic Kondo systems cannot be explained by the spectral functions
exhibiting a bias-dependent split Kondo peak which is observed by a three-terminal setup$^{17}$.
Our secondary purpose is to understand the spin dynamics forming each coherent peak. The spin dynamics
of a two-reservoir mesoscopic Kondo system is obtained from the eigenvector of the Liouvillian matrix
instead of the Hamiltonian matrix. The former matrix has been given in our previous study$^{18}$.
In contrast, the Hamiltonian matrix cannot be easily obtained.

We begin by discussing the dynamics of EKS tunnelling under bias.
The spin exchange in a singlet is the only coherent dynamics in a
conventional single-reservoir Kondo system at equilibrium.
However, an additional coherent dynamics plays an important role in a two-reservoir mesoscopic Kondo system.
We depict such new dynamics, i.e., singlet hopping, in Fig.~1b. The next step of Fig.~1b can be
divided into two types: one including the spin exchange process and the other excluding it.
The transition from Fig.~1b to Fig.~1c, which comprises singlet hopping, spin exchange,
and singlet partner change in order, is an example of the former type, while the latter type,
from Fig.~1b to Fig.~1d, is a successive singlet hopping with singlet partner change.
Singlet partner changing is not a particle dynamics but an important quantum process
in a coherent state. The former tunnelling, denoted as EKS-1, occurs
near zero bias because spin exchange is the major dynamics in equilibrium.
In contrast, the latter tunnelling type, denoted as EKS-2, occurs at a finite bias.
The co-tunnelling without spin exchange, i.e., EKS-2, is more effective than the EKS-1 in
forming coherent current. We analytically clarify these properties of EKS tunnelling
in the later part of this study.

The important property of EKS tunnelling under bias is moving together unidirectionally.
This property of EKS gives rise to the following three aspects:
(i) neglecting multiple spin exchange process,
(ii) making local density of states (LDOS) bias-independent unless a quasiparticle
is excited by bias, and (iii) excluding linear response regime. These three aspects are crucial
in understanding the transport in a mesoscopic Kondo system under bias.


The tunnelling conductance at zero temperature is written as
\begin{equation}\label{eq1}
 dI/dV=(e/2\hbar) {\widetilde
\Gamma}[\rho^{ss}_{m}(eV/2)+\rho^{ss}_{m}(-eV/2)],
\end{equation}
where ${\widetilde\Gamma}=\Gamma^L\Gamma^R/(\Gamma^L+
\Gamma^R)$ for the flat density of states of the left ($L$) and right ($R$)
metallic reservoirs and $\rho^{ss}_{m}(\omega)$,
where $ss$ denotes the steady-state nonequilibrium, is the LDOS at
the mediating Kondo atom. Equation (\ref{eq1}) is derived from the well-known Meir-Wingreen
current formula$^{19,20}$ using the proportionate coupling function
condition, $\Gamma^L\propto\Gamma^R$, and the bias independence of the LDOS.
In other studies on Kondo-involved mesoscopic systems$^{21-23}$,
a bias-independent $\rho^{ss}_{m}(\omega)$ has also been adopted.
The LDOS $\rho_{m\uparrow}^{ss}(\omega)$ is given by $\rho_{m\uparrow}^{ss}(\omega)=(1/\pi){\rm
Re}[({\bf M})^{-1}]_{mm}$, where ${\bf M}$ is the Liouvillian matrix provided in the Methods section.
We employ the two-reservoir Anderson impurity model:
\begin{equation}\label{eq4}
{\cal H}={\cal H}_0^{L,R}+{\cal H}^{L,R}_C+\sum_{\sigma}\epsilon_mc^\dagger_{m\sigma} c_{m\sigma}
+Un_{m\uparrow}n_{m\downarrow},
\end{equation}
where ${\cal H}_0^{L,R}=\sum_{k,\sigma}(\epsilon_k-\mu^{L,R})c^{\dagger}_{k\sigma}
c_{k\sigma}$, ${\cal H}_C^{L,R}=\sum_{k,\sigma}(V_{km}^{L,R}c^\dagger
_{m\sigma}c_{k\sigma}+V^{L,R*}_{km}c^{\dagger}_{k\sigma}
c_{m\sigma})$, and $\sigma$, $\epsilon_{k}$, $\epsilon_{m}$,
$V_{km}$, $U$, and $\mu$ indicate the electron spin, kinetic
energy, energy level of the mediating atom, hybridization
strength, on-site Coulomb repulsion, and chemical potential,
respectively, to study the two-reservoir mesoscopic Kondo system under bias.
We set $V_{km}^{L,R}=V$ in this study.

We have shown in a previous study$^{24}$ that neglecting the basis vectors
describing multiple spin exchange is a good approximation to extract the core Kondo
dynamics in a single-reservoir Kondo system at equilibrium. In an entangled two-reservoir system,
the unidirectional motion of EKS further validates
neglecting multiple spin exchange. Thus, we construct a working
Liouville space$^{18,24}$ valid in the large-$U$ regime, which is spanned by a complete
set $\{{\hat e}_p\}$ comprising \\
$\{c_{k\uparrow}^L, \, \, \delta
n_{m\downarrow}c_{k\uparrow}^L\}$, $k=0, 1, 2, \cdots, \infty$, \\
$\{\delta j^{+L}_{m\downarrow}c_{m\uparrow}, \, \, \delta
j^{-L}_{m\downarrow}c_{m\uparrow}, \, \, c_{m\uparrow}, \, \,
\delta j^{-R}_{m\downarrow}c_{m\uparrow}, \, \, \delta
j^{+R}_{m\downarrow}c_{m\uparrow}\}$, and \\ $\{\delta
n_{m\downarrow}c_{k\uparrow}^R, \, \,  c_{k\uparrow}^R\}$,
$k=0, 1, 2, \cdots, \infty$, \\
where $j^-_{m\downarrow}=i\sum_k(V_{km}c^\dagger_{m\downarrow}c_{k\downarrow}-
V^*_{km}c^\dagger_{k\downarrow}c_{m\downarrow})$,
$j^+_{m\downarrow}=\sum_k(V_{km}c^\dagger_{m\downarrow}c_{k\downarrow}
+V^*_{km}c^\dagger_{k\downarrow}c_{m\downarrow})$, and $\delta$ indicates $\delta A=A-\langle
A\rangle$. The angular bracket denotes the average. The $\delta$ is
introduced to achieve orthogonality among the basis vectors.
For convenience, we omit the normalization factors $\langle(\delta
j^{\pm L,R}_{m\downarrow})^2\rangle^{1/2}$ and $\langle(\delta
n_{m\downarrow})^2\rangle^{1/2}$ in the denominators of the
corresponding basis vectors. Thus, the fluctuations are naturally involved.

We arrange the basis vectors in the order given above
and construct the matrix ${\rm\bf M}$. One can reduce the
infinite-dimensional matrix ${\rm\bf M}$ to a $5\times 5$
matrix ${\rm\bf M}^{r}_{5\times 5}$ by using the matrix partitioning technique$^{25}$.
Then, the LDOS is given by $\rho_{m\uparrow}^{ss}(\omega)=
(1/\pi){\rm Re}[({\bf M}^r_{5\times 5})^{-1}]_{33}$, where
\begin{eqnarray} {\rm\bf M}^{r}_{5\times 5}
=\left( \begin{array}{c c c c c} -i\omega' & \gamma_{_{LL}} &
-U^L_{j^-} & \gamma_{_{LR}} & \gamma_{_{j^-}} \\ -\gamma_{_{LL}} &
-i\omega' &
 -U^L_{j^+} & \gamma_{_{j^+}} & \gamma_{_{LR}} \\
U_{j^-}^{L*} &  U_{j^+}^{L*} & -i\omega' &  U^{R*}_{j^+} &
U^{R*}_{j^-} \\  -\gamma_{_{LR}} & -\gamma_{_{j^+}} & -U_{j^+}^R &
-i\omega' & -\gamma_{_{RR}} \\
 -\gamma_{_{j^-}} &  -\gamma_{_{LR}} &
 -U_{j^-}^R  & \gamma_{_{RR}} &  -i\omega'
\end{array} \right), \label{r2r}
\end{eqnarray}
$\omega'\equiv\omega-\epsilon_m-U\langle
n_{m\downarrow}\rangle$, and $\langle n_{m\downarrow}\rangle$
denotes the average number of down-spin electrons occupying the
mediating atom.
In equation (\ref{r2r}), all the matrix elements, except for $U_{j^\pm}^{L,R}$,
have additional self-energy terms $-i\Sigma_{pq}=-\beta_{pq}[i\Sigma^L_0(\omega)+i\Sigma^R_0(\omega)]$,
where $\Sigma^{L(R)}_{0}(\omega)=-i\Gamma^{L(R)}/2$ denotes the self-energy of
${\cal H}_0^{L(R)}$ for a flat wide band. The coefficients $\beta_{pq}$ appear in the process of matrix
reduction$^{18}$. We use $\Delta\equiv(\Gamma^L+\Gamma^R)/4$ as an energy unit.

The matrix ${\rm\bf M}^{r}_{5\times 5}$ in equation (\ref{r2r}) consists of two
$3\times 3$ blocks that share the central element representing the
mediating atom and two $2\times 2$ blocks at the corners.
The three off-diagonal elements of the $3\times 3$ block represent
the degrees of singlet coupling $\gamma_{_{LL(RR)}}=\langle\widehat{V}[j^{-L(R)}_{m\downarrow},j^{+L(R)}_{m\downarrow}]\rangle$,
where $\widehat{V}=\sum_kiV(c_{k\uparrow}^L+c_{k\uparrow}^R)c^\dagger_{m\uparrow}$,
and the incoherent double occupancy parameters $U_{j^\pm}^{L,R}$, whose explicit form is given in ref.~18.

The $2\times 2$ corner blocks describe inter-reservoir coherence$^{16}$. The matrix elements are written as
$\gamma_{_{LR}}=\langle \widehat{V}\sum_{r\in R}\sum_{l\in
L}|V|^2(c^\dagger_{l\downarrow}c_{r\downarrow}
+c^\dagger_{r\downarrow}c_{l\downarrow})\rangle$ and
$\gamma_{_{j}}=\gamma_{_{j^\mp}}=\langle \widehat{V}
\sum_{r\in R}\sum_{l\in L}|V|^2(c^\dagger_{l\downarrow}c_{r\downarrow}
-c^\dagger_{r\downarrow}c_{l\downarrow})\rangle$, representing successive back and forth double hopping.
The negative sign in the antidiagonal element $\gamma_{_{j}}$ indicates that it represents the effect of bias, whereas the element $\gamma_{_{LR}}$ having positive sign represents inter-reservoir coherent dynamics, which is linearly independent of the $\gamma_{_{j}}$ dynamics. The unidirectional motion of EKS gives rise to the condition $\gamma_{_{j}}=\gamma_{_{LR}}$ when a bias is applied.

Now, we reproduce the experimental $dI/dV$ line shapes.  Specifically, we choose
the line shape of a QPC with the closest side peaks given in Fig.~1(b) of ref.~8 and scanning tunnelling
spectroscopy (STS) for a Co atom placed on a Cu$_2$N layer covering a Cu (100) substrate
in ref.~9. These systems explicitly show both the zero-bias peak and the two side peaks.
For the QPC, we employ the spontaneous formation scenario of a localized spin at
the bound state$^{26,27}$. Therefore, the Hamiltonian for the two-reservoir
Anderson impurity model, equation (2), is commonly applicable to these systems.

We obtain the line shapes in terms of the matrix elements provided in Table I and the
coefficients $\beta_{pq}$ given below:
${\rm Re}[\beta_{12}]={\rm Re}[\beta_{14}]={\rm Re}[\beta_{15}]=0.247$, ${\rm Re}[\beta_{11}]
={\rm Re}[\beta_{22}]={\rm Re}[\beta_{44}]={\rm
Re}[\beta_{55}]={\rm Re}[\beta_{24}]={\rm Re}[\beta_{25}]={\rm
Re}[\beta_{45}]=0.253$. ${\rm Re}[\beta_{33}]$ is exactly unity. These values are chosen
based on the standard value $0.25$, which is the value at the atomic limit$^{18}$.
The difference among ${\rm Re}[\beta_{pq}]$ is attributable to the different signs
for the current, i.e., $\langle j^{-L}_{m\downarrow}\rangle=-\langle
j^{-R}_{m\downarrow}\rangle<0$.
We superimpose our theoretical results on the experimental data in Figs.~2a and 2b using
the energy units $\Delta=$ 0.5 meV for Fig.~2a and $\Delta=$ 3 meV for Fig.~2b.
The fittings are remarkable. Choi {\it et al}.$^{10}$, who studied the same system
shown in the inset of Fig.~2b, observed two side peaks which appear as shoulders in Fig.~2b.
Therefore, the theoretical $dI/dV$ line shapes given in Figs.~2a and 2b explicitly illustrate
the two coherent side peaks, which are attributed to inter-reservoir coherence$^{16}$.

\begin{figure}
[t] \vspace*{6.0cm} \includegraphics{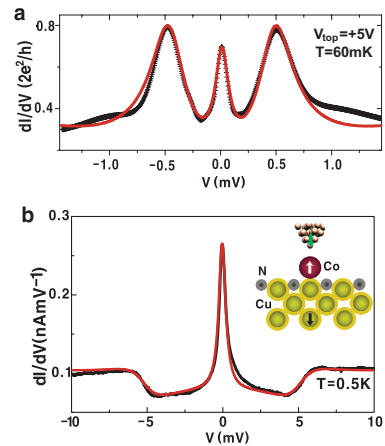} \vspace*{0.0cm}
\caption{\textbf{Comparisons between theory (red) and experiment (black).}
{\bf a,} Experimental data of the closest side peaks reported in ref.~8. We choose
$\widetilde{\Gamma}/\Delta=0.83$ to obtain theoretical curve. {\bf b,} STS line shape of ref.~9.
We use arbitrary units for theoretical curve. Inset shows the STS
setup.}
\end{figure}

The values of the matrix elements are determined phenomenologically
because our operator formalism does not directly determine the many-body parameters.
For the left-right symmetric system such as the QPC of Fig.~2a, we choose
${\rm Re}U_{j^+}^{L}={\rm Re}U_{j^+}^{R}$. Also, we consider that $j^+$ contribution is
larger than $j^-$ contribution and adopt inequality
${\rm Re}U_{j^+}^{L,R}>{\rm Re}U_{j^-}^{L,R}$. Hence, we choose the values, as shown in Table I-2a.
The STS setup of Fig.~2b is left (substrate)-right (tip) asymmetric, and we adopt
the inequalities ${\rm Re}U_{j^+}^{L}>{\rm Re}U_{j^-}^{L,R}>{\rm Re}U_{j^+}^{R}$ for Fig.~2b, from which
the values in Table I-2b are chosen. In this STS measurement, the insulating layer plays
an important role by which the degrees of Kondo couplings are enhanced.
The imaginary parts of $\beta_{pq}$ and $U_{j^\pm}^{L,R}$ vanish at half-filling$^{18}$,
which is the case in this study.


\begin{table}[b]
Table I: \textbf{Matrix elements for Figs.~2 and 3.}

\begin{tabular}{c c c c c c c c}
\hline\hline \,  Fig. \, & \, $\gamma_{_{LL}}$ \, & $\gamma_{_{RR}}$ \,
& $\gamma_{_{j,LR}}$ \, & $
{\rm Re}U^{L,R}_{j^-}$ & ${\rm Re}U_{j^+}^{L}$ & ${\rm
Re}U_{j^+}^{R}$ & ${\rm Im}U_{j^\pm}^{L,R}$
\\ [0.5ex] \hline
2a & 0.5  & 0.5 &  0.65 & 1.05 & 1.26 & 1.26  & 0  \\
2b & 0.8  & 0.7 &  0.43 & 2.8 & 7.0 & 1.62  & 0  \\
 3 & 0.008  & 0.008 &  0.15 & 0.52 & 0.525 & 0.525  & 0  \\
[1ex] \hline
\end{tabular} \label{table1}
\end{table}


\begin{figure}
[t] \vspace*{3cm} \includegraphics{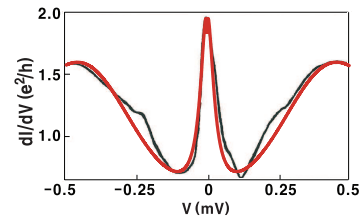} \vspace*{0.0cm}
\caption{\textbf{Comparison of QDSET line shapes.}
The theoretical result (red) is superposed on the experimental data (black)
for a QDSET reported in ref.~2. }
\end{figure}

Now, we obtain the $dI/dV$ line shape of a QDSET in the unitary regime$^2$.
The line shape structure of Fig.~3 looks quite different from those two in Fig.~2
because two coherent side peaks are not explicitly shown. The broad side peaks
are enhanced Coulomb peaks. The QDSET line shape is obtained by
adopting relatively weak Kondo couplings, as shown in Table I-3. Weak Kondo coupling
results in the side peaks being very close to each other and
causes large fluctuations in the denominator of ${\rm Re}[\beta_{pq}]$, which
enhances the Coulomb peaks. Because of left-right symmetry of QDSET, we adopt ${\rm Re}U_{j^+}^{L}={\rm Re}U_{j^+}^{R}$ and ${\rm Re}U_{j^+}^{S,R}>{\rm Re}U_{j^-}^{L,R}$ like the QPC, as shown in Table I-3.
We choose ${\rm Re}[\beta_{pq}]$ to be $0.45$ times
the previous ${\rm Re}[\beta_{pq}]$ and fit the experiment using the energy unit $\Delta=$ 0.11 meV.
The small offset observed in the field-induced peak splitting shown in refs.~5 and 6 demonstrates
that the two side peaks are near each other, and the peak at higher voltage becomes the major peak.

Lastly, we reveal the dynamics of EKS tunnelling from the eigenvectors of the Liouvillian matrix $iL$, which
is given by using $\omega^\prime=0$ in equation (\ref{r2r}). We study the dynamics at the atomic limit,
i.e., $U_{j^\pm}^{L,R}=U/4$ and $\Sigma_{pq}=0$ and use $\gamma_{_{LL}}=\gamma_{_{RR}}=\gamma$
for simplicity. The steady-state condition $\gamma_{_j}=\gamma_{_{LR}}$ is definitely used.
Then, the Liouvillian matrix in the large-$U$ regime has the eigenvalues
$0$ (EKS-1), $\pm i\gamma$ (EKS-2), and $\pm iU/2$ and the corresponding eigenvectors
$[1, -1, 0, -1, 1]^T$ (EKS-1), $[-1, \pm i, 0, \mp i, 1]^T$ (EKS-2), and
$[\pm 1, \pm 1, 0, \pm 1, \pm 1]^T$, respectively. The superscript $T$ denotes the transpose.
The channel at $\omega^\prime=-\gamma$, which corresponds to the eigenvalue $i\gamma$, is for hole tunnelling.
Each element of the eigenvector $[1, -1, 0, -1, 1]^T$, for example, corresponds to the basis operator
$j^{-L}_{m\downarrow}c_{m\uparrow}$, $-j^{+L}_{m\downarrow}c_{m\uparrow}$, $c_{m\uparrow}$,
$-j^{+R}_{m\downarrow}c_{m\uparrow}$, and $j^{-R}_{m\downarrow}c_{m\uparrow}$, respectively.
The single reservoir Anderson model, which involves only the spin exchange process as coherent dynamics,
has an eigenvector $[0, -1, 1]^T$ with respect to the eigenvalue $0$.
This indicates that the left three elements or the right three elements in $[1, -1, 0, -1, 1]^T$
describe the spin exchange process. In contrast, those three elements in $[-1, \pm i, 0, \mp i, 1]^T$
indicate that the spin exchange process is prohibited in the EKS-2 tunnelling. On the other hand, the
singlet hopping, which is allowed only in a two-reservoir Anderson model, is described by the $(1,5)$
and $(2,4)$ pairs. Interestingly, the eigenvectors show that the EKS-1 has an
$L-R$ symmetric singlet hopping, whereas the EKS-2 has an $L-R$ antisymmetric singlet hopping.
This fact is consistent with the double-well problem, in which
the antisymmetric combination has a higher energy state than the symmetric combination.

In summary, we reveal that the two-reservoir mesoscopic Kondo system has two different coherent transport channels, i.e., EKS-1 and EKS-2. The former is responsible for the zero-bias and the latter for the two side peaks.
Our results fit the experimental $dI/dV$ line shapes remarkably well.
We also reveal from the Liouvillian matrix that the fundamental difference between EKS-1 and EKS-2 originates from different symmetry in establishing entanglement. Our Liouvillian analysis may be extended to explaining the tunnelling conductance for a sample with strong electron correlation.

\noindent{\bf Methods}

We use the operator formalism in the Liouville space in which the LDOS at the mediating atom $\rho_{m\uparrow}^{ss}(\omega)$
is written as $\rho_{m\uparrow}^{ss}(\omega)=-(1/\pi){\rm
Im}G^{+}_{mm\uparrow}(\omega)$, where $G^{+}_{mm\uparrow}(\omega)=\langle
c_{m\uparrow}|(\omega-{\rm\bf L})^{-1}|c_{m\uparrow}\rangle$. ${\rm\bf
L}$ is the Liouville operator defined as ${\rm\bf L}A={\cal H}A-A{\cal H}$,
where ${\cal H}$ is the Hamiltonian and $A$ is an operator. The inner product is defined by $\langle A|B\rangle\equiv\langle AB^\dagger+B^\dagger A\rangle$, where $B^\dagger$ is the adjoint
of $B$. Hence, we write $\rho_{m\uparrow}^{ss}(\omega)=(1/\pi){\rm
Re}[({\bf M})^{-1}]_{mm}$, where the matrix elements of ${\bf M}$
are given by ${\rm\bf M}_{pq}=-i\omega\delta_{pq}+\langle{\hat
e}_q|i{\rm\bf L}{\hat e}_p\rangle$. The working Liouville space is spanned by
a complete set $\{{\hat e}_p\}$ provided in the text.

JH thanks P. Coleman for suggesting the entangled Kondo singlet and  S.-W. Cheong,
P. Kim, A. Millis, and P. Fulde for valuable discussions.
This research was supported by the Basic Science Research Program through the NRF, Korea
(2012R1A1A2005220), and was partially supported by a KIAS grant funded by MEST.



\end{document}